\documentclass[aps,prl,twocolumn,noshowpacs,superscriptaddress,groupedaddress]{revtex4}  % for review and submission
\usepackage{graphicx}  % needed for figures
\usepackage{dcolumn}   % needed for some tables
\usepackage{bm}        % for math
\usepackage{amssymb}   % for math
\usepackage{romannum}
\usepackage{wasysym}
\usepackage{color}
\definecolor{mygray}{gray}{0.3}

\begin{document}

\title{A Zeeman slower for diatomic molecules}
\author{M. Petzold, P. Kaebert, P. Gersema, M. Siercke, S. Ospelkaus}

\affiliation{Leibniz Universit\"at Hannover, Institute for Quantum Optics,Welfengarten 1, D-30167 Hannover, Germany}

\date{\today}

\begin{abstract}
We present a novel slowing scheme for beams of laser-coolable diatomic molecules reminiscent of Zeeman slowing of atomic beams. The scheme results in efficient compression of the 1-dimensional velocity distribution to velocities trappable by magnetic or magneto-optical traps.  3D Monte Carlo simulations for the prototype molecule $^{88}\mathrm{Sr}^{19}\mathrm{F}$ and experiments in an atomic testbed demonstrate a performance comparable to traditional atomic Zeeman slowing and an enhancement of flux below v=35 m/s by a factor of $\approx 20$ compared to white-light slowing. This is the first experimentally shown continuous and dissipative slowing technique in molecule-like level structures, promising to provide the missing link for the preparation of large ultracold molecular ensembles.
\end{abstract}
\maketitle

Cooling molecular ensembles to temperatures near absolute zero has been a goal of the ultracold community for decades. Such ultracold ensembles would enable research on new phases of matter, precision measurements and ultracold chemistry \cite{carr_cold_2009}. Current research on direct laser cooling of molecules with quasi-diagonal Franck-Condon structure has had much success, demonstrating magnetic and magneto-optical traps \cite{truppe_molecules_2017,anderegg_radio_2017,barry_magneto-optical_2014,hummon_2d_2013,mccarron_magnetically-trapped_2017,williams_magnetic_2017} and optical molasses \cite{truppe_molecules_2017,kozyryev_sisyphus_2017,lim_ultracold_2017,mccarron_magnetically-trapped_2017,williams_magnetic_2017}, reaching temperatures of $\approx 50\, \mathrm{\mu K}$ \cite{truppe_molecules_2017}. Their ultimate success in producing large ultracold ensembles, however, is currently severely hampered by the lack of an efficient %, continuous and high flux 
source of slow molecules at velocities trappable in magnetic or magneto-optical traps.

While a variety of slowing methods for rovibrationally cold molecular beams exist, including two stage buffer gas cooling \cite{lu_cold_2011}, Stark and Zeeman deceleration \cite{meerakker_stark_2006,narevicius_stopping_2008}, centrifuge deceleration \cite{chervenkov_continuous_2014}, white-light slowing \cite{barry_laser_2012,hemmerling_laser_2016} and chirped light slowing \cite{truppe_intense_2017,yeo_rotational_2015},  all shown techniques are either not continuous, have only poor control on the final velocity or do not compress the 1-dimensional velocity distribution of the molecules. 
This limits the number of molecules loaded into magnetic or magneto-optical traps to a fraction of the numbers in atomic experiments \cite{anderegg_radio_2017,truppe_molecules_2017, norrgard_submillikelvin_2016, lu_magnetic_2014}.  Zeeman slowing, the only technique combining all the aforementioned advantages, was up to now considered to be impossible to implement for laser-coolable molecules \cite{barry_laser_2012,yeo_rotational_2015}. 

Here we present for the first time a Zeeman slower scheme for laser-coolable molecules, resulting in  continuous deceleration and compression of the molecular velocity distribution down to velocities in the 10 m/s range. In the following, we shortly review the  traditional atomic Zeeman slower concept, discuss problems arising from the complex molecular level structure and show how these problems can be overcome with our molecular Zeeman slowing concept. We  perform 3D Monte Carlo simulations of the scheme for the prototype molecule $\mathrm{^{88}Sr^{19}F}$ and finally  implement our scheme in an atomic testbed.

\begin{center}
\begin{figure*}
\onecolumngrid
	\includegraphics[width=1\textwidth]{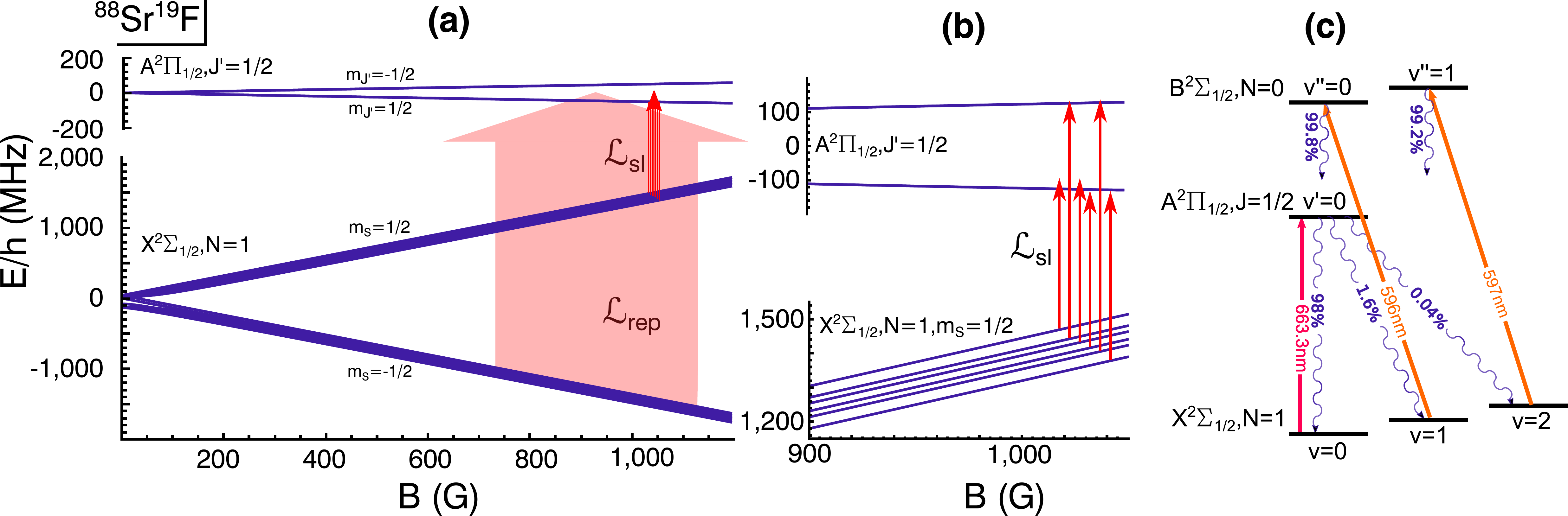}
		\caption{\label{Figure1} (a) Type \Romannum{2} Zeeman slower scheme on the $X^2\Sigma_{1/2},N=1, \nu=0 \rightarrow A^2\Pi_{1/2},J'=1/2, \nu'=0$ transition. As long as $g_{_{\Pi}} \ll g_{_{\Sigma}}$ the level structure in high magnetic fields resembles an effective 3-level system. The graph is shown for the prototype molecule $\mathrm{^{88}Sr^{19}F}$. (b) Necessary sidebands on the slowing laser $\mathcal{L}_{sl}$. (c) Proposed vibrational repumping scheme for the prototype molecule $\mathrm{^{88}Sr^{19}F}$. Branching ratios are taken from \cite{barry_magneto-optical_2014, ramachandran_franck-condon_2005}. } 
\end{figure*}
\end{center}
 A traditional atomic Zeeman slower \cite{phillips_laser_1982} works on a type \Romannum{1} level structure, where the angular momentum of the excited state $J'=J+1$ is larger than that of the ground state $J$. 
Here the atoms get pumped into a bright, stretched state at which point they cycle in an effective 2-level system \cite{phillips_laser_1982}, whose transition frequency is tunable by a magnetic field. The atomic beam is then radiatively slowed down by a counter-propagating laser beam while an inhomogeneous magnetic field compensates the changing Doppler shift during the slowing process down to a well defined final velocity. In analogy to magneto-optical traps (MOTs) working on a type \Romannum{1} level structure (typically referred to as type \Romannum{1} MOT) we will refer to this as type \Romannum{1} Zeeman slowing.

In contrast, laser-coolable molecules require working in a type \Romannum{2} level structure ($J \rightarrow J'=J$ or $J \rightarrow J'=J-1$) to prohibit decay into higher lying rotational states \cite{stuhl_magneto-optical_2008}. As a natural consequence, molecules are optically pumped into dark magnetic sublevels rather than bright ones and a 2-level cycling transition does not exist. Nevertheless, slowing and cooling of molecules using laser light has been realized by destabilizing these dark states \cite{berkeland_destabilization_2002}. As a result, all magnetic sublevels of the ground state need to be coupled to the excited state, each of which will exhibit a different shift in energy in a magnetic field, thus preventing the implementation of a traditional Zeeman slower in these type \Romannum{2} systems.

We propose a solution to this problem for laser-coolable molecular radicals working on the $X^2\Sigma_{1/2},N=1, \nu=0$ (ground state described in Hund's case b) to $A^2\Pi_{1/2},J'=1/2, \nu'=0$ (excited state described in Hund's case a) transition, where $\nu,\nu'$ are the respective vibrational quantum numbers, $N$ is the rotational angular momentum in the ground state and $J'$ is the total angular momentum in the excited state. By adding a large magnetic offset field $B_{0}$ to the traditional Zeeman slower design, the electron spin decouples from the nuclear and rotational angular momenta in the ground state, splitting it into two manifolds with $m_S=\pm1/2$ (see Fig. \ref{Figure1} a).
 The sublevels inside these respective manifolds are shifted equally in energy with increasing or decreasing magnetic field strength. The excited state splits into $m_J'=\pm1/2$ manifolds with a much smaller splitting due to a smaller g-factor $g_{_{\Pi}}\ll g_{_{\Sigma}} $. In the limit of negligible hyperfine structure and vanishing $g_{_{\Pi}}$, this reduces to an effective 3-level system (see Fig. \ref{Figure1} a).

To implement a type \Romannum{2} Zeeman slower in this 3-level system, the $m_S=1/2$ manifold of the ground state is coupled to the excited state via a narrow linewidth (on the order of the transition linewidth), counter-propagating laser beam at saturation intensity. This transition is magnetically tunable and therefore can be used to compensate for a changing Doppler shift during the slowing process, as it is done in traditional type \Romannum{1} Zeeman slowing. In the following, we will refer to this laser beam as the "slowing laser" $\mathcal{L}_{sl}$. 
Due to the large spin orbit coupling in the excited state, molecules can decay back to either $m_S=1/2$ or to $m_S=-1/2$. A frequency broadened laser (in the following referred to as the "repumping laser" $\mathcal{L}_{rep}$) pumps molecules at all relevant velocities and magnetic fields from $m_S=-1/2$ back to the slowing transition. 
Molecules traveling fast enough to see $\mathcal{L}_{sl}$ on resonance due to the Doppler shift, get pumped between the $m_{S}=\pm1/2$ manifolds by scattering photons from $\mathcal{L}_{sl}$ and $\mathcal{L}_{rep}$ until they are shifted out of resonance with $\mathcal{L}_{sl}$.
%Coherent dark states are destabilized due to the broadness of L2. 
Further slowing of the molecules occurs with changing magnetic field, bringing the molecules back into resonance with $\mathcal{L}_{sl}$. Since slower molecules feel no force while faster ones are being slowed down, we achieve both compression of the velocity distribution and reduction of the mean molecular velocity by spatially varying the magnetic field.

In a realistic system, including finite hyperfine structure of the ground state as well as a small upper state g-factor $g_{_{\Pi}}$, the slowing laser $\mathcal{L}_{sl}$ needs to couple every hyperfine state in the $m_S=+1/2$ ground state manifold to the excited state as shown in Fig. \ref{Figure1} b). This can be realised by a suitable choice of sideband frequencies. Our scheme is applicable to all laser-coolable molecules, where $g_{_{\Pi}} \ll g_{_{\Sigma}}$, so that the simplified 3-level picture holds in the Paschen-Back Regime including for  example  $\mathrm{^{88}Sr^{19}F},g_{_{\Pi}}\approx -0.08$, $\mathrm{CaF},g_{_{\Pi}}\approx -0.02$ and $\mathrm{YO},g_{_{\Pi}}\approx -0.06$.

To go beyond the qualitative discussion of a three-level system and demonstrate the feasibility of the scheme in a realistic system including hyperfine structure,  we  now focus on the prototype molecule $\mathrm{^{88}Sr^{19}F}$ with nuclear spin $I=1/2$ \footnote{Detailed simulations for $\mathrm{^{40}Ca^{19}F}$ will be discussed in a later publication.}. Fig. \ref{Figure1} a) shows a plot of the $\mathrm{^{88}Sr^{19}F}$ $X^2\Sigma_{1/2},N=1, \nu=0 \rightarrow A^2\Pi_{1/2},J'=1/2, \nu'=0$ level structure as a function of magnetic field. The groundstate manifolds $m_S=\pm 1/2$ each split into 6 sublevels ($m_{N}=\pm1,0;m_{I}=\pm1/2$) (due to rotational and hyperfine structure), which have to be coupled  to the 4 sublevels of the $A^2\Pi_{1/2}, J'=1/2, \nu'=0$ state ($m_{J}=\pm1/2;m_{I}=\pm1/2$)  via the slowing laser $\mathcal{L}_{sl}$ as shown in Fig. \ref{Figure1} b). This specific system thus requires a slowing laser $\mathcal{L}_{sl}$ with 6 sidebands. Pairs of frequencies of $\mathcal{L}_{sl}$ that couple to the same excited state are detuned by $\delta=\pm \Gamma/2$ from resonance (where $\Gamma\approx 2 \pi\times6.6 \, \mathrm{MHz}$ is the transition linewidth) to avoid pumping into coherent dark states. Furthermore, a broad repumper $\mathcal{L}_{rep}$ with a width of  $\Delta f \approx 1.1 \, \mathrm{GHz}$ is required to pump $m_S=-1/2$ molecules back into the cooling cycle.
 To calculate the velocity dependent force profile  along the slowing path, we solve the 16-level optical Bloch equations at  magnetic offset fields of $B=900\, \mathrm{G}$,$B=1000\, \mathrm{G}$ and $B=1050\, \mathrm{G}$ respectively    \footnote{ Transition rates and level energies are calculated by diagonalizing the Hamiltonian (including rotational and hyperfine structure) and compared to rates obtained from Pgopher \cite{western_pgopher:_2017}. The calculation further assumes that laser beams contain $\sigma^+$ and $\sigma^-$ polarized light, with no $\pi$ polarization (i.e. the laser beams travel along the background magnetic field).}. Each frequency in the slowing laser $\mathcal{L}_{sl}$ is assumed to have an intensity of $24 \, \mathrm{mW \, cm^{-2}}$ corresponding to a Rabi frequency of $\Omega_{ij}=2 \Gamma d_{ij}$, where $d_{ij}$ is the normalized dipole matrix element of the respective transition. Coupling from the $m_S=-1/2$ states with the repump laser $\mathcal{L}_{rep}$ is modeled by an electric field, frequency modulated at $\omega_{mod}=\pi \Gamma$ with a modulation index of $27$ and an intensity of $860\, \mathrm{mW \, cm^{-2}}$ corresponding to $\Omega_{ij}=12 \Gamma d_{ij}$. Due to the modulation, no additional dark states arise from $\mathcal{L}_{rep}$. The calculation results in a narrow velocity-dependent  force profile, which can be tuned over the whole relevant velocity range by a spatially varying magnetic field, consistent with the idea of Zeeman slowing (see Fig. 2).
 Note that loss of molecules during the cooling cycle  due to  vibrational branching is largely suppressed due to the quasi-diagonal Franck-Condon structure of molecular radicals and can furthermore be suppressed by  an experimentally feasible repumping scheme  via the $B^2\Sigma_{1/2}, N=0,\nu'=0,1$ state (see Fig. 1 c)).

\begin{figure}
	\includegraphics[width=0.47\textwidth]{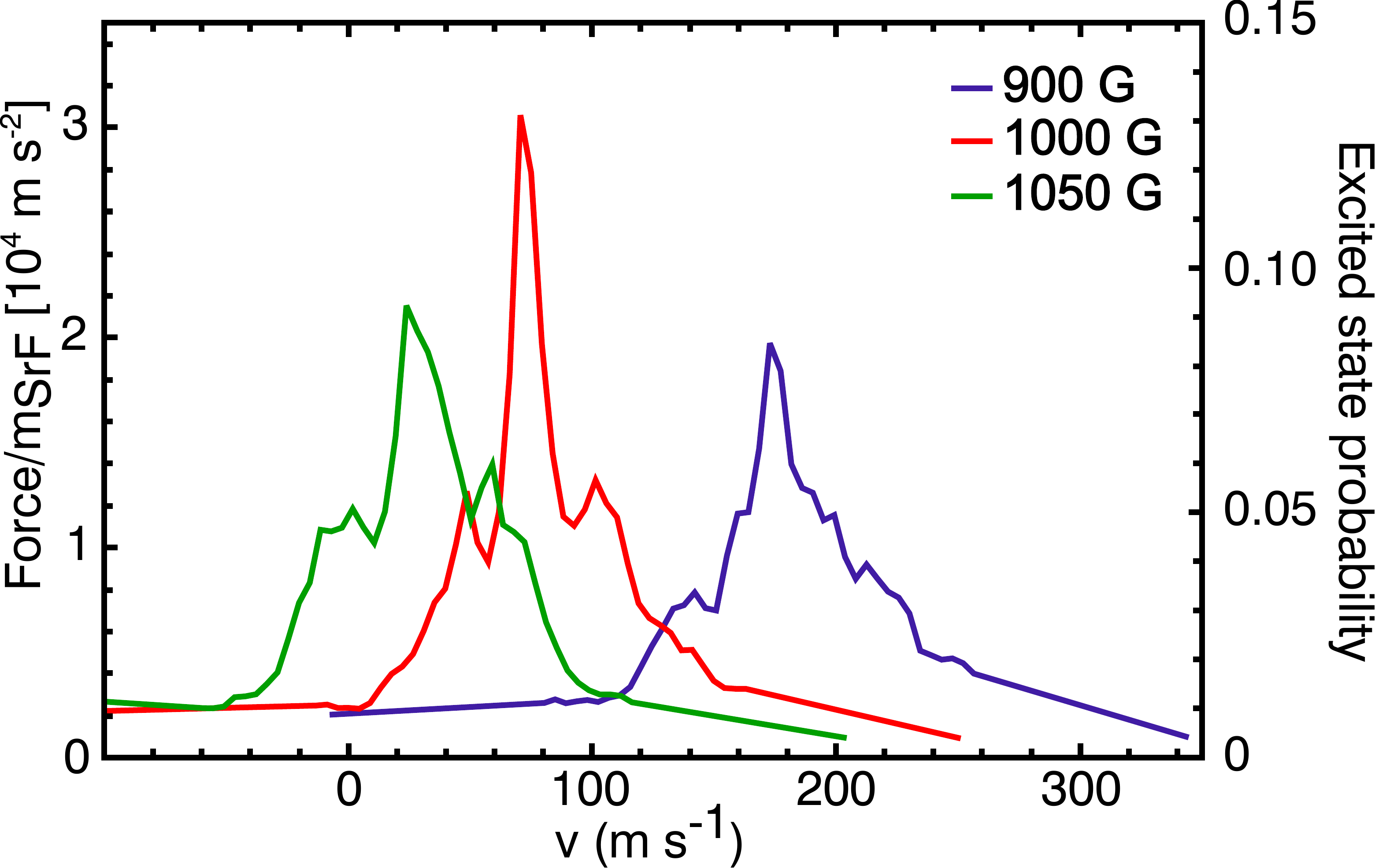}
	\caption{\label{Figure2} Velocity dependent $\mathrm{^{88}Sr^{19}F}$ Zeeman slower force profiles and corresponding excited state probabilities at different magnetic fields (for details see text). The substructure in the profiles emerges due to the hyperfine structure of the groundstate and the finite excited state $g_{_{\Pi}}$.}
	
\end{figure}

\begin{figure}
	\includegraphics[width=0.43\textwidth]{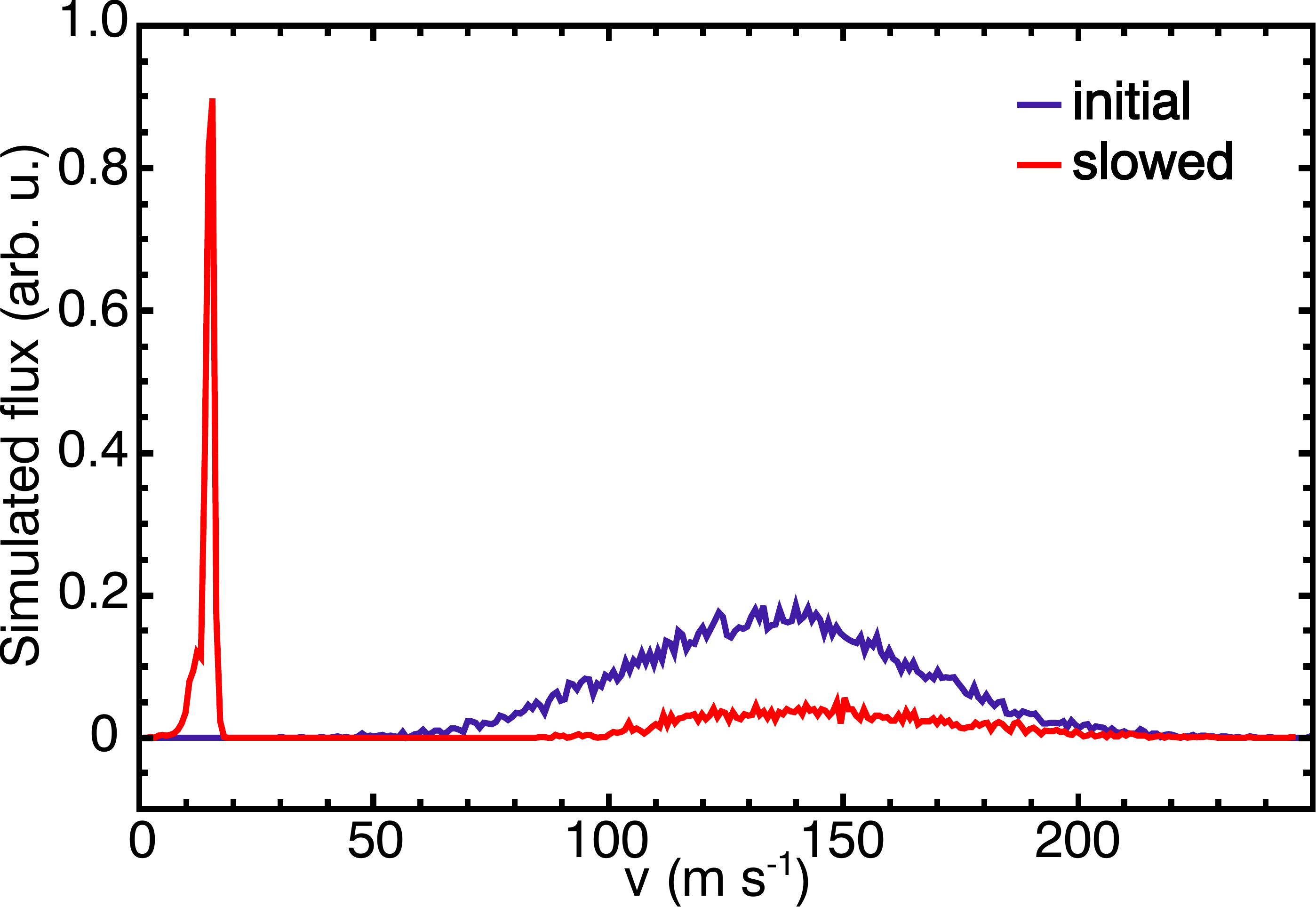}%
\caption{\label{Figure3} 3D Monte Carlo simulation of Zeeman slowing for a $\mathrm{^{88}Sr^{19}F}$ molecular beam originating from a buffer gas cell. Blue: Initial velocity distribution detected at $z_{det}=1.58 \mathrm{\, m}$ behind the buffer gas cell. Red: Slowed distribution based on the force profile from Fig. \ref{Figure2}. $20 \, \%$ of the molecules are within the peak centered around $v_{end}=15 \, \mathrm{m} \, \mathrm{s}^{-1}$. Further details about the simulation can be found in the text.}\end{figure}

We now use the force profile from Fig. 2 in a 3D Monte Carlo simulation to calculate the velocity profile of Zeeman slowed $\mathrm{^{88}Sr^{19}F}$ molecules originating from a typical cryogenic buffer gas cell with an initial longitudinal velocity distribution  centered at $v_{l}=120   \, \mathrm{m \, s^{-1}}$,  a longitudinal velocity spread of $\Delta v_{l}= 75 \, \mathrm{m \, s^{-1}}$ at full width half maximum and a transverse velocity spread of $\Delta v_{t}= 80 \, \mathrm{m \, s^{-1}}$ (see Fig. \ref{Figure3}). We assume the magnetic field to rise from $B_{0}=\mathrm{900 \, G}$ at $z=\mathrm{0.35 \, m}$ behind the buffer gas cell to $B_{max}=1030 \, \mathrm{G}$ at $z=\mathrm{1.33 \, \mathrm{m}}$. Furthermore we divide the force profile in Fig. \ref{Figure2} by a safety factor of 2 and take transverse heating effects during the slowing process into account. The detection region is chosen to be located at $z_{\mathrm{det}}=1.58 \, \mathrm{m}$ behind the exit of the buffer gas cell and is restricted to a (R x L) $\mathrm{0.3 \, cm}$ x $\mathrm{3 \, cm}$ cylinder. This geometry corresponds to the experimental setup of our demonstration experiment described below. As can be seen in Fig. 3, the simulation results in a significant fraction ($20 \, \%$) of the initial molecules to be slowed over the entire Zeeman slower path and  compressed to a velocity distribution centered at $v_{end}=15 \, \mathrm{m} \, \mathrm{s}^{-1}$ with a full width at half maximum of $\Delta v_{l}\approx2.5 \, \mathrm{m} \, \mathrm{s}^{-1}$.  Due to the combination of slowing and compression characteristic for a Zeeman slower,  a significant fraction of molecules exits the Zeeman slower with velocities low enough to be efficiently captured with existing trapping schemes. 

\begin{center}
\begin{figure*}
\onecolumngrid
	\includegraphics[width=0.95\textwidth]{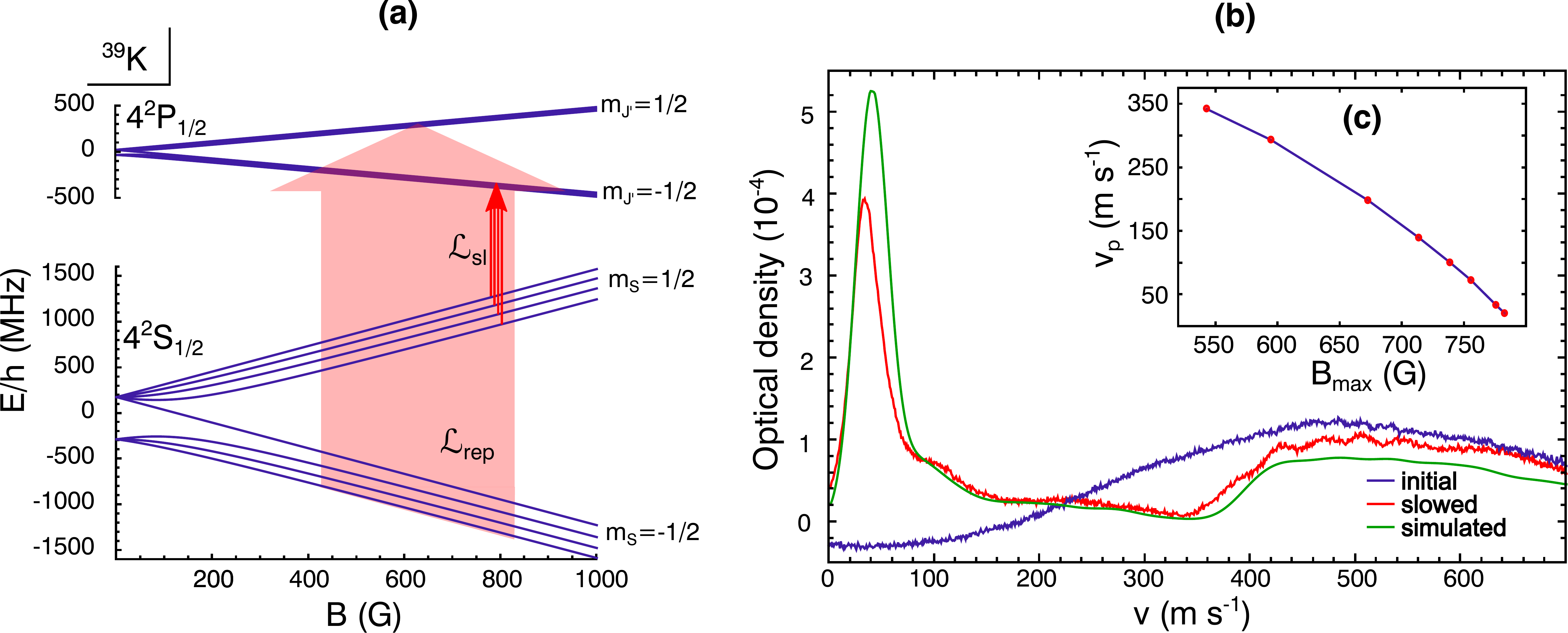}
	\caption{\label{Figure4} Experimental demonstration of type \Romannum{2} Zeeman slowing on the $\mathrm{^{39}K \, D_{1}}$- line: \textbf{(a)} Potassium level structure for the $4^2S_{1/2}$ ground state and $4^2P_{1/2}$ excited state in the Paschen-Back regime. \textbf{(b)} Measured velocity distributions for a beam of $^{39}$K atoms before (blue) and after type \Romannum{2} Zeeman slowing (red). Comparison to the Monte Carlo simulation (green). \textbf{(c)} The final peak velocity of the atomic beam $v_{p}$ as a function of the magnetic field maximum $B_{max}$ at the end of the Zeeman slowing region.}
\end{figure*}
\end{center}
Ultimately, we demonstrate   type \Romannum{2} Zeeman slowing  in an atomic testbed on a transition comparable to the  $X^2\Sigma_{1/2},N=1, \nu=0 \rightarrow A^2\Pi_{1/2},J'=1/2, \nu'=0$ transition of a molecular radical. For this purpose, we pick the  $\mathrm{D_{1}}$-line of $\mathrm{^{39}K}$ atoms, a $J\rightarrow J'=J$ transition showing striking similarity to the  $\mathrm{^{88}Sr^{19}F}$ transition discussed above (compare Fig. \ref{Figure4} a) and Fig. \ref{Figure1} a). 
In our experiment, we make use of an atomic beam source  with a peak velocity at $450 \, \mathrm{m\,s^{-1}}$.  We apply a $B_{0}=510 \, \mathrm{G}$ magnetic offset field in the $130 \, \mathrm{cm}$ long slowing region to bring the potassium atoms in the Paschen-Back regime. At this offset field,  the slowing laser $\mathcal{L}_{sl}$ consists of 4 frequencies each $118.6\, \mathrm{MHz}$ apart to couple the transitions $\mathrm{4^{2}S_{1/2}}\left| m_{J}=1/2,m_{I}=-3/2,...,3/2 \right\rangle \rightarrow\mathrm{4^{2}P_{1/2}}\left| m_{J'}=-1/2,m_{I}=-3/2,...,3/2\right\rangle $ respectively and is locked $1680\, \mathrm{MHz}$ red of the $\mathrm{D_{1}}$-line crossover of a Doppler free potassium spectroscopy. The repumping laser $\mathcal{L}_{rep}$ is frequency-broadened to an approximate width of $1.5\, \mathrm{GHz}$ through current modulation of a free running DFB-diode with modulation frequency of $12\, \mathrm{MHz}$  to pump atoms from the $m_{S}=-1/2$ manifold back to the slowing cycle at all relevant magnetic fields and velocities.
  Throughout the slowing region the magnetic field is increased from $B_{0}=510\, \mathrm{G}$ to $B_{max}=770\, \mathrm{G}$ corresponding to a capture velocity of $v_{cap}=400\, \mathrm{m}\, \mathrm{s}^{-1}$ and a expected final velocity of $v_{end}=35\, \mathrm{m}\, \mathrm{s}^{-1}$ at the end of the slowing region. We probe the longitudinal velocity distribution $25\, \mathrm{cm}$ behind the end of the slowing region with differential absorption Doppler spectroscopy, %. By using absorption spectroscopy
 where we detect atoms in a region which is restricted by the detection beam diameter $\mathrm{d_{beam}}=3 \, \mathrm{mm}$ times the diameter of the vacuum tube $\mathrm{d_{tube}}=3 \, \mathrm{cm}$.

 Fig. \ref{Figure4} b shows the experimentally measured velocity profile after type \Romannum{2} Zeeman slowing along with a simulated velocity profile fed by the experimentally expected Maxwell Boltzmann distribution originating from an oven running at T=450~K.
 
 The measured and simulated profiles   show deceleration and compression of the 1-dimensional velocity distribution of the atomic beam. The final peak velocity $v_{p}$ is easily tunable through the current in the magnetic field coil and the corresponding magnetic field maximum (see Fig. \ref{Figure4} c)). Small differences between simulation and experiment may be due to non perfect beam-overlap, photon recoil of the detection laser, non perfect background subtraction, stray magnetic fields in the detection region or non perfect spectral distribution of the repumping laser. The good overall agreement of the simulation with the experiment is a strong argument for the validity of the $\mathrm{^{88}Sr^{19}F}$ simulation shown in Fig. \ref{Figure3}.
The efficiency of our method can be quantified by comparing the measured flux of atoms below $v_{end}=35\, \mathrm{m\, s^{-1}}$ of $\Phi_{\mathrm{type \Romannum{2}}}=3.3*10^{9} \mathrm{\frac{atm}{cm^{2} \, s}}$ to that of a type \Romannum{1} traditional atomic Zeeman slower working on the $\mathrm{D_{2}}$-line of $\mathrm{^{39}K}$ and to white-light slowing on the $\mathrm{D_{1}}$-line in the same setup. 
We find  the type \Romannum{2} Zeeman slower to reach nearly the same performance as the well established type \Romannum{1} traditional atomic Zeeman slower $\Phi_{\mathrm{type \Romannum{2}}} / \Phi_{\mathrm{type \Romannum{1}}}=0.6$. Moreover, the type \Romannum{2} Zeeman slower outperforms white-light slowing, the current standard technique for molecular beam slowing, by a factor of $\Phi_{\mathrm{type \Romannum{2}}} / \Phi_{\mathrm{white}}=20$. A detailed comparison between these slowing methods is beyond the scope of this proposal and will follow in a later publication.

Type \Romannum{2} Zeeman slowing should be applicable to most of today's laser-coolable molecules with realistic experimental requirements, as long as $g_{_{\Pi}} \ll g_{_{\Sigma}}$, including the presented case of $\mathrm{^{88}Sr^{19}F},g_{_{\Pi}}\approx -0.08$ as well as the already laser-cooled species $\mathrm{CaF},g_{_{\Pi}}\approx -0.02$ and $\mathrm{YO},g_{_{\Pi}}\approx -0.06$.  $\mathrm{BaF}$ with $g_{_{\Pi}}\approx -0.2$ might need a slightly frequency broadened slowing laser for implementation.  As $\mathcal{L}_{sl}$ and $\mathcal{L}_{rep}$ are both far detuned from resonance in low magnetic fields, the slowing scheme is ideally suited to be continuously coupled to already existing trapping schemes without disturbing already trapped molecules. Because of the continuous nature of Zeeman slowing, it can ideally be combined with current pulsed molecular buffer gas sources for the realization of a quasi-continous loading scheme by loading a whole sequence of pulses or with continuous sources of rovibrationally cold molecules instead of pulsed ones in future experiments.  This will increase the flux of molecules even further, opening the possibility to realize large magneto-optical traps as an efficient starting point for work towards molecular Bose-Einstein condensates and quantum degenerate Fermi gases with exciting prospects for applications including precision measurements, ultracold chemistry and dipolar quantum many-body systems.

This work was supported by the Centre for Quantum Engineering and SpaceTime Research QUEST. M.P. acknowledges financial  support from the Deutsche Forschungsgemeinschaft through Research Training Group 1729, and M.S. through Research Training Group 1991.


\begin{thebibliography}{20}
\expandafter\ifx\csname natexlab\endcsname\relax\def\natexlab#1{#1}\fi
\expandafter\ifx\csname bibnamefont\endcsname\relax
  \def\bibnamefont#1{#1}\fi
\expandafter\ifx\csname bibfnamefont\endcsname\relax
  \def\bibfnamefont#1{#1}\fi
\expandafter\ifx\csname citenamefont\endcsname\relax
  \def\citenamefont#1{#1}\fi
\expandafter\ifx\csname url\endcsname\relax
  \def\url#1{\texttt{#1}}\fi
\expandafter\ifx\csname urlprefix\endcsname\relax\def\urlprefix{URL }\fi
\providecommand{\bibinfo}[2]{#2}
\providecommand{\eprint}[2][]{\url{#2}}

\bibitem[{\citenamefont{Carr et~al.}(2009)\citenamefont{Carr, DeMille, Krems,
  and Ye}}]{carr_cold_2009}
\bibinfo{author}{\bibfnamefont{L.~D.} \bibnamefont{Carr}},
  \bibinfo{author}{\bibfnamefont{D.}~\bibnamefont{DeMille}},
  \bibinfo{author}{\bibfnamefont{R.~V.} \bibnamefont{Krems}}, \bibnamefont{and}
  \bibinfo{author}{\bibfnamefont{J.}~\bibnamefont{Ye}}, \bibinfo{journal}{New
  Journal of Physics} \textbf{\bibinfo{volume}{11}}, \bibinfo{pages}{055049}
  (\bibinfo{year}{2009}),
  \urlprefix\url{http://stacks.iop.org/1367-2630/11/i=5/a=055049}.

\bibitem[{\citenamefont{Truppe et~al.}(2017{\natexlab{a}})\citenamefont{Truppe,
  Williams, Hambach, Caldwell, Fitch, Hinds, Sauer, and
  Tarbutt}}]{truppe_molecules_2017}
\bibinfo{author}{\bibfnamefont{S.}~\bibnamefont{Truppe}},
  \bibinfo{author}{\bibfnamefont{H.~J.} \bibnamefont{Williams}},
  \bibinfo{author}{\bibfnamefont{M.}~\bibnamefont{Hambach}},
  \bibinfo{author}{\bibfnamefont{L.}~\bibnamefont{Caldwell}},
  \bibinfo{author}{\bibfnamefont{N.~J.} \bibnamefont{Fitch}},
  \bibinfo{author}{\bibfnamefont{E.~A.} \bibnamefont{Hinds}},
  \bibinfo{author}{\bibfnamefont{B.~E.} \bibnamefont{Sauer}}, \bibnamefont{and}
  \bibinfo{author}{\bibfnamefont{M.~R.} \bibnamefont{Tarbutt}},
  \bibinfo{journal}{Nat Phys} \textbf{\bibinfo{volume}{advance online
  publication}} (\bibinfo{year}{2017}{\natexlab{a}}), ISSN
  \bibinfo{issn}{1745-2481},
  \urlprefix\url{http://dx.doi.org/10.1038/nphys4241}.

\bibitem[{\citenamefont{Anderegg et~al.}(2017)\citenamefont{Anderegg,
  Augenbraun, Chae, Hemmerling, Hutzler, Ravi, Collopy, Ye, Ketterle, and
  Doyle}}]{anderegg_radio_2017}
\bibinfo{author}{\bibfnamefont{L.}~\bibnamefont{Anderegg}},
  \bibinfo{author}{\bibfnamefont{B.~L.} \bibnamefont{Augenbraun}},
  \bibinfo{author}{\bibfnamefont{E.}~\bibnamefont{Chae}},
  \bibinfo{author}{\bibfnamefont{B.}~\bibnamefont{Hemmerling}},
  \bibinfo{author}{\bibfnamefont{N.~R.} \bibnamefont{Hutzler}},
  \bibinfo{author}{\bibfnamefont{A.}~\bibnamefont{Ravi}},
  \bibinfo{author}{\bibfnamefont{A.}~\bibnamefont{Collopy}},
  \bibinfo{author}{\bibfnamefont{J.}~\bibnamefont{Ye}},
  \bibinfo{author}{\bibfnamefont{W.}~\bibnamefont{Ketterle}}, \bibnamefont{and}
  \bibinfo{author}{\bibfnamefont{J.~M.} \bibnamefont{Doyle}},
  \bibinfo{journal}{Phys. Rev. Lett.} \textbf{\bibinfo{volume}{119}},
  \bibinfo{pages}{103201} (\bibinfo{year}{2017}),
  \urlprefix\url{https://link.aps.org/doi/10.1103/PhysRevLett.119.103201}.

\bibitem[{\citenamefont{Barry et~al.}(2014)\citenamefont{Barry, McCarron,
  Norrgard, Steinecker, and DeMille}}]{barry_magneto-optical_2014}
\bibinfo{author}{\bibfnamefont{J.~F.} \bibnamefont{Barry}},
  \bibinfo{author}{\bibfnamefont{D.~J.} \bibnamefont{McCarron}},
  \bibinfo{author}{\bibfnamefont{E.~B.} \bibnamefont{Norrgard}},
  \bibinfo{author}{\bibfnamefont{M.~H.} \bibnamefont{Steinecker}},
  \bibnamefont{and} \bibinfo{author}{\bibfnamefont{D.}~\bibnamefont{DeMille}},
  \bibinfo{journal}{Nature} \textbf{\bibinfo{volume}{512}},
  \bibinfo{pages}{286} (\bibinfo{year}{2014}), ISSN \bibinfo{issn}{0028-0836},
  \urlprefix\url{http://dx.doi.org/10.1038/nature13634}.

\bibitem[{\citenamefont{Hummon et~al.}(2013)\citenamefont{Hummon, Yeo, Stuhl,
  Collopy, Xia, and Ye}}]{hummon_2d_2013}
\bibinfo{author}{\bibfnamefont{M.~T.} \bibnamefont{Hummon}},
  \bibinfo{author}{\bibfnamefont{M.}~\bibnamefont{Yeo}},
  \bibinfo{author}{\bibfnamefont{B.~K.} \bibnamefont{Stuhl}},
  \bibinfo{author}{\bibfnamefont{A.~L.} \bibnamefont{Collopy}},
  \bibinfo{author}{\bibfnamefont{Y.}~\bibnamefont{Xia}}, \bibnamefont{and}
  \bibinfo{author}{\bibfnamefont{J.}~\bibnamefont{Ye}}, \bibinfo{journal}{Phys.
  Rev. Lett.} \textbf{\bibinfo{volume}{110}}, \bibinfo{pages}{143001}
  (\bibinfo{year}{2013}),
  \urlprefix\url{https://link.aps.org/doi/10.1103/PhysRevLett.110.143001}.

\bibitem[{\citenamefont{McCarron et~al.}(2017)\citenamefont{McCarron,
  Steinecker, Zhu, and DeMille}}]{mccarron_magnetically-trapped_2017}
\bibinfo{author}{\bibfnamefont{D.~J.} \bibnamefont{McCarron}},
  \bibinfo{author}{\bibfnamefont{M.~H.} \bibnamefont{Steinecker}},
  \bibinfo{author}{\bibfnamefont{Y.}~\bibnamefont{Zhu}}, \bibnamefont{and}
  \bibinfo{author}{\bibfnamefont{D.}~\bibnamefont{DeMille}},
  \bibinfo{journal}{ArXiv e-prints}  (\bibinfo{year}{2017}).

\bibitem[{\citenamefont{Williams et~al.}(2017)\citenamefont{Williams, Caldwell,
  Fitch, Truppe, Rodewald, Hinds, Sauer, and Tarbutt}}]{williams_magnetic_2017}
\bibinfo{author}{\bibfnamefont{H.~J.} \bibnamefont{Williams}},
  \bibinfo{author}{\bibfnamefont{L.}~\bibnamefont{Caldwell}},
  \bibinfo{author}{\bibfnamefont{N.~J.} \bibnamefont{Fitch}},
  \bibinfo{author}{\bibfnamefont{S.}~\bibnamefont{Truppe}},
  \bibinfo{author}{\bibfnamefont{J.}~\bibnamefont{Rodewald}},
  \bibinfo{author}{\bibfnamefont{E.~A.} \bibnamefont{Hinds}},
  \bibinfo{author}{\bibfnamefont{B.~E.} \bibnamefont{Sauer}}, \bibnamefont{and}
  \bibinfo{author}{\bibfnamefont{M.~R.} \bibnamefont{Tarbutt}},
  \bibinfo{journal}{ArXiv e-prints}  (\bibinfo{year}{2017}).

\bibitem[{\citenamefont{Lim et~al.}(2017)\citenamefont{Lim, Almond, Trigatzis,
  Devlin, Fitch, Sauer, Tarbutt, and Hinds}}]{lim_ultracold_2017}
\bibinfo{author}{\bibfnamefont{J.}~\bibnamefont{Lim}},
  \bibinfo{author}{\bibfnamefont{J.~R.} \bibnamefont{Almond}},
  \bibinfo{author}{\bibfnamefont{M.~A.} \bibnamefont{Trigatzis}},
  \bibinfo{author}{\bibfnamefont{J.~A.} \bibnamefont{Devlin}},
  \bibinfo{author}{\bibfnamefont{N.~J.} \bibnamefont{Fitch}},
  \bibinfo{author}{\bibfnamefont{B.~E.} \bibnamefont{Sauer}},
  \bibinfo{author}{\bibfnamefont{M.~R.} \bibnamefont{Tarbutt}},
  \bibnamefont{and} \bibinfo{author}{\bibfnamefont{E.~A.} \bibnamefont{Hinds}},
  \bibinfo{journal}{ArXiv e-prints}  (\bibinfo{year}{2017}).



\bibitem[{\citenamefont{Ramachandran et~al.}(2005)\citenamefont{Ramachandran,
  Rajamanickam, Bagare, and Kumar}}]{ramachandran_franck-condon_2005}
\bibinfo{author}{\bibfnamefont{P.~S.} \bibnamefont{Ramachandran}},
  \bibinfo{author}{\bibfnamefont{N.}~\bibnamefont{Rajamanickam}},
  \bibinfo{author}{\bibfnamefont{S.~P.} \bibnamefont{Bagare}},
  \bibnamefont{and} \bibinfo{author}{\bibfnamefont{B.~C.} \bibnamefont{Kumar}},
  \bibinfo{journal}{Astrophysics and Space Science}
  \textbf{\bibinfo{volume}{295}}, \bibinfo{pages}{443} (\bibinfo{year}{2005}),
  ISSN \bibinfo{issn}{1572-946X},
  \urlprefix\url{https://doi.org/10.1007/s10509-005-0743-4}.


\bibitem[{\citenamefont{Kozyryev et~al.}(2017)\citenamefont{Kozyryev, Baum,
  Matsuda, Augenbraun, Anderegg, Sedlack, and Doyle}}]{kozyryev_sisyphus_2017}
\bibinfo{author}{\bibfnamefont{I.}~\bibnamefont{Kozyryev}},
  \bibinfo{author}{\bibfnamefont{L.}~\bibnamefont{Baum}},
  \bibinfo{author}{\bibfnamefont{K.}~\bibnamefont{Matsuda}},
  \bibinfo{author}{\bibfnamefont{B.~L.} \bibnamefont{Augenbraun}},
  \bibinfo{author}{\bibfnamefont{L.}~\bibnamefont{Anderegg}},
  \bibinfo{author}{\bibfnamefont{A.~P.} \bibnamefont{Sedlack}},
  \bibnamefont{and} \bibinfo{author}{\bibfnamefont{J.~M.} \bibnamefont{Doyle}},
  \bibinfo{journal}{Phys. Rev. Lett.} \textbf{\bibinfo{volume}{118}},
  \bibinfo{pages}{173201} (\bibinfo{year}{2017}),
  \urlprefix\url{https://link.aps.org/doi/10.1103/PhysRevLett.118.173201}.

\bibitem[{\citenamefont{Lu et~al.}(2011)\citenamefont{Lu, Rasmussen, Wright,
  Patterson, and Doyle}}]{lu_cold_2011}
\bibinfo{author}{\bibfnamefont{H.-I.} \bibnamefont{Lu}},
  \bibinfo{author}{\bibfnamefont{J.}~\bibnamefont{Rasmussen}},
  \bibinfo{author}{\bibfnamefont{M.~J.} \bibnamefont{Wright}},
  \bibinfo{author}{\bibfnamefont{D.}~\bibnamefont{Patterson}},
  \bibnamefont{and} \bibinfo{author}{\bibfnamefont{J.~M.} \bibnamefont{Doyle}},
  \bibinfo{journal}{Phys. Chem. Chem. Phys.} \textbf{\bibinfo{volume}{13}},
  \bibinfo{pages}{18986} (\bibinfo{year}{2011}),
  \urlprefix\url{http://dx.doi.org/10.1039/C1CP21206K}.

\bibitem[{\citenamefont{Meerakker et~al.}(2006)\citenamefont{Meerakker,
  Vanhaecke, and Meijer}}]{meerakker_stark_2006}
\bibinfo{author}{\bibfnamefont{S.~Y. T. v.~d.} \bibnamefont{Meerakker}},
  \bibinfo{author}{\bibfnamefont{N.}~\bibnamefont{Vanhaecke}},
  \bibnamefont{and} \bibinfo{author}{\bibfnamefont{G.}~\bibnamefont{Meijer}},
  \bibinfo{journal}{Annual Review of Physical Chemistry}
  \textbf{\bibinfo{volume}{57}}, \bibinfo{pages}{159} (\bibinfo{year}{2006}),
  \urlprefix\url{http://dx.doi.org/10.1146/annurev.physchem.55.091602.094337}.

\bibitem[{\citenamefont{Narevicius et~al.}(2008)\citenamefont{Narevicius,
  Libson, Parthey, Chavez, Narevicius, Even, and
  Raizen}}]{narevicius_stopping_2008}
\bibinfo{author}{\bibfnamefont{E.}~\bibnamefont{Narevicius}},
  \bibinfo{author}{\bibfnamefont{A.}~\bibnamefont{Libson}},
  \bibinfo{author}{\bibfnamefont{C.~G.} \bibnamefont{Parthey}},
  \bibinfo{author}{\bibfnamefont{I.}~\bibnamefont{Chavez}},
  \bibinfo{author}{\bibfnamefont{J.}~\bibnamefont{Narevicius}},
  \bibinfo{author}{\bibfnamefont{U.}~\bibnamefont{Even}}, \bibnamefont{and}
  \bibinfo{author}{\bibfnamefont{M.~G.} \bibnamefont{Raizen}},
  \bibinfo{journal}{Phys. Rev. Lett.} \textbf{\bibinfo{volume}{100}},
  \bibinfo{pages}{093003} (\bibinfo{year}{2008}),
  \urlprefix\url{https://link.aps.org/doi/10.1103/PhysRevLett.100.093003}.

\bibitem[{\citenamefont{Chervenkov et~al.}(2014)\citenamefont{Chervenkov, Wu,
  Bayerl, Rohlfes, Gantner, Zeppenfeld, and
  Rempe}}]{chervenkov_continuous_2014}
\bibinfo{author}{\bibfnamefont{S.}~\bibnamefont{Chervenkov}},
  \bibinfo{author}{\bibfnamefont{X.}~\bibnamefont{Wu}},
  \bibinfo{author}{\bibfnamefont{J.}~\bibnamefont{Bayerl}},
  \bibinfo{author}{\bibfnamefont{A.}~\bibnamefont{Rohlfes}},
  \bibinfo{author}{\bibfnamefont{T.}~\bibnamefont{Gantner}},
  \bibinfo{author}{\bibfnamefont{M.}~\bibnamefont{Zeppenfeld}},
  \bibnamefont{and} \bibinfo{author}{\bibfnamefont{G.}~\bibnamefont{Rempe}},
  \bibinfo{journal}{Phys. Rev. Lett.} \textbf{\bibinfo{volume}{112}},
  \bibinfo{pages}{013001} (\bibinfo{year}{2014}),
  \urlprefix\url{https://link.aps.org/doi/10.1103/PhysRevLett.112.013001}.

\bibitem[{\citenamefont{Barry et~al.}(2012)\citenamefont{Barry, Shuman,
  Norrgard, and DeMille}}]{barry_laser_2012}
\bibinfo{author}{\bibfnamefont{J.~F.} \bibnamefont{Barry}},
  \bibinfo{author}{\bibfnamefont{E.~S.} \bibnamefont{Shuman}},
  \bibinfo{author}{\bibfnamefont{E.~B.} \bibnamefont{Norrgard}},
  \bibnamefont{and} \bibinfo{author}{\bibfnamefont{D.}~\bibnamefont{DeMille}},
  \bibinfo{journal}{Phys. Rev. Lett.} \textbf{\bibinfo{volume}{108}},
  \bibinfo{pages}{103002} (\bibinfo{year}{2012}),
  \urlprefix\url{https://link.aps.org/doi/10.1103/PhysRevLett.108.103002}.

\bibitem[{\citenamefont{Hemmerling et~al.}(2016)\citenamefont{Hemmerling, Chae,
  Ravi, Anderegg, Drayna, Hutzler, Collopy, Ye, Ketterle, and
  Doyle}}]{hemmerling_laser_2016}
\bibinfo{author}{\bibfnamefont{B.}~\bibnamefont{Hemmerling}},
  \bibinfo{author}{\bibfnamefont{E.}~\bibnamefont{Chae}},
  \bibinfo{author}{\bibfnamefont{A.}~\bibnamefont{Ravi}},
  \bibinfo{author}{\bibfnamefont{L.}~\bibnamefont{Anderegg}},
  \bibinfo{author}{\bibfnamefont{G.~K.} \bibnamefont{Drayna}},
  \bibinfo{author}{\bibfnamefont{N.~R.} \bibnamefont{Hutzler}},
  \bibinfo{author}{\bibfnamefont{A.~L.} \bibnamefont{Collopy}},
  \bibinfo{author}{\bibfnamefont{J.}~\bibnamefont{Ye}},
  \bibinfo{author}{\bibfnamefont{W.}~\bibnamefont{Ketterle}}, \bibnamefont{and}
  \bibinfo{author}{\bibfnamefont{J.~M.} \bibnamefont{Doyle}},
  \bibinfo{journal}{Journal of Physics B: Atomic, Molecular and Optical
  Physics} \textbf{\bibinfo{volume}{49}}, \bibinfo{pages}{174001}
  (\bibinfo{year}{2016}),
  \urlprefix\url{http://stacks.iop.org/0953-4075/49/i=17/a=174001}.

\bibitem[{\citenamefont{Truppe et~al.}(2017{\natexlab{b}})\citenamefont{Truppe,
  Williams, Fitch, Hambach, Wall, Hinds, Sauer, and
  Tarbutt}}]{truppe_intense_2017}
\bibinfo{author}{\bibfnamefont{S.}~\bibnamefont{Truppe}},
  \bibinfo{author}{\bibfnamefont{H.~J.} \bibnamefont{Williams}},
  \bibinfo{author}{\bibfnamefont{N.~J.} \bibnamefont{Fitch}},
  \bibinfo{author}{\bibfnamefont{M.}~\bibnamefont{Hambach}},
  \bibinfo{author}{\bibfnamefont{T.~E.} \bibnamefont{Wall}},
  \bibinfo{author}{\bibfnamefont{E.~A.} \bibnamefont{Hinds}},
  \bibinfo{author}{\bibfnamefont{B.~E.} \bibnamefont{Sauer}}, \bibnamefont{and}
  \bibinfo{author}{\bibfnamefont{M.~R.} \bibnamefont{Tarbutt}},
  \bibinfo{journal}{New Journal of Physics} \textbf{\bibinfo{volume}{19}},
  \bibinfo{pages}{022001} (\bibinfo{year}{2017}{\natexlab{b}}),
  \urlprefix\url{http://stacks.iop.org/1367-2630/19/i=2/a=022001}.

\bibitem[{\citenamefont{Yeo et~al.}(2015)\citenamefont{Yeo, Hummon, Collopy,
  Yan, Hemmerling, Chae, Doyle, and Ye}}]{yeo_rotational_2015}
\bibinfo{author}{\bibfnamefont{M.}~\bibnamefont{Yeo}},
  \bibinfo{author}{\bibfnamefont{M.~T.} \bibnamefont{Hummon}},
  \bibinfo{author}{\bibfnamefont{A.~L.} \bibnamefont{Collopy}},
  \bibinfo{author}{\bibfnamefont{B.}~\bibnamefont{Yan}},
  \bibinfo{author}{\bibfnamefont{B.}~\bibnamefont{Hemmerling}},
  \bibinfo{author}{\bibfnamefont{E.}~\bibnamefont{Chae}},
  \bibinfo{author}{\bibfnamefont{J.~M.} \bibnamefont{Doyle}}, \bibnamefont{and}
  \bibinfo{author}{\bibfnamefont{J.}~\bibnamefont{Ye}}, \bibinfo{journal}{Phys.
  Rev. Lett.} \textbf{\bibinfo{volume}{114}}, \bibinfo{pages}{223003}
  (\bibinfo{year}{2015}),
  \urlprefix\url{https://link.aps.org/doi/10.1103/PhysRevLett.114.223003}.

\bibitem[{\citenamefont{Norrgard et~al.}(2016)\citenamefont{Norrgard, McCarron,
  Steinecker, Tarbutt, and DeMille}}]{norrgard_submillikelvin_2016}
\bibinfo{author}{\bibfnamefont{E.~B.} \bibnamefont{Norrgard}},
  \bibinfo{author}{\bibfnamefont{D.~J.} \bibnamefont{McCarron}},
  \bibinfo{author}{\bibfnamefont{M.~H.} \bibnamefont{Steinecker}},
  \bibinfo{author}{\bibfnamefont{M.~R.} \bibnamefont{Tarbutt}},
  \bibnamefont{and} \bibinfo{author}{\bibfnamefont{D.}~\bibnamefont{DeMille}},
  \bibinfo{journal}{Phys. Rev. Lett.} \textbf{\bibinfo{volume}{116}},
  \bibinfo{pages}{063004} (\bibinfo{year}{2016}),
  \urlprefix\url{https://link.aps.org/doi/10.1103/PhysRevLett.116.063004}.

\bibitem[{\citenamefont{Lu et~al.}(2014)\citenamefont{Lu, Kozyryev, Hemmerling,
  Piskorski, and Doyle}}]{lu_magnetic_2014}
\bibinfo{author}{\bibfnamefont{Hsin-I.}~\bibnamefont{Lu}},
  \bibinfo{author}{\bibfnamefont{I.}~\bibnamefont{Kozyryev}},
  \bibinfo{author}{\bibfnamefont{B.}~\bibnamefont{Hemmerling}},
  \bibinfo{author}{\bibfnamefont{J.}~\bibnamefont{Piskorski}},
  \bibnamefont{and} \bibinfo{author}{\bibfnamefont{J.~M.} \bibnamefont{Doyle}},
  \bibinfo{journal}{Phys. Rev. Lett.} \textbf{\bibinfo{volume}{112}},
  \bibinfo{pages}{113006} (\bibinfo{year}{2014}),
  \urlprefix\url{https://link.aps.org/doi/10.1103/PhysRevLett.112.113006}.

\bibitem[{\citenamefont{Phillips and Metcalf}(1982)}]{phillips_laser_1982}
\bibinfo{author}{\bibfnamefont{W.~D.} \bibnamefont{Phillips}} \bibnamefont{and}
  \bibinfo{author}{\bibfnamefont{H.}~\bibnamefont{Metcalf}},
  \bibinfo{journal}{Phys. Rev. Lett.} \textbf{\bibinfo{volume}{48}},
  \bibinfo{pages}{596} (\bibinfo{year}{1982}),
  \urlprefix\url{https://link.aps.org/doi/10.1103/PhysRevLett.48.596}.

\bibitem[{\citenamefont{Stuhl et~al.}(2008)\citenamefont{Stuhl, Sawyer, Wang,
  and Ye}}]{stuhl_magneto-optical_2008}
\bibinfo{author}{\bibfnamefont{B.~K.} \bibnamefont{Stuhl}},
  \bibinfo{author}{\bibfnamefont{B.~C.} \bibnamefont{Sawyer}},
  \bibinfo{author}{\bibfnamefont{D.}~\bibnamefont{Wang}}, \bibnamefont{and}
  \bibinfo{author}{\bibfnamefont{J.}~\bibnamefont{Ye}}, \bibinfo{journal}{Phys.
  Rev. Lett.} \textbf{\bibinfo{volume}{101}}, \bibinfo{pages}{243002}
  (\bibinfo{year}{2008}),
  \urlprefix\url{https://link.aps.org/doi/10.1103/PhysRevLett.101.243002}.

\bibitem[{\citenamefont{Berkeland and
  Boshier}(2002)}]{berkeland_destabilization_2002}
\bibinfo{author}{\bibfnamefont{D.~J.} \bibnamefont{Berkeland}}
  \bibnamefont{and} \bibinfo{author}{\bibfnamefont{M.~G.}
  \bibnamefont{Boshier}}, \bibinfo{journal}{Phys. Rev. A}
  \textbf{\bibinfo{volume}{65}}, \bibinfo{pages}{033413}
  (\bibinfo{year}{2002}),
  \urlprefix\url{https://link.aps.org/doi/10.1103/PhysRevA.65.033413}.

\bibitem[{\citenamefont{Western}(2017)}]{western_pgopher:_2017}
\bibinfo{author}{\bibfnamefont{C.~M.} \bibnamefont{Western}},
  \bibinfo{journal}{Journal of Quantitative Spectroscopy and Radiative
  Transfer} \textbf{\bibinfo{volume}{186}}, \bibinfo{pages}{221 }
  (\bibinfo{year}{2017}), ISSN \bibinfo{issn}{0022-4073},
  \urlprefix\url{http://www.sciencedirect.com/science/article/pii/S0022407316300437}.

\end{thebibliography}
\end{document}